\documentclass[preprint]{imsart}

\RequirePackage[OT1]{fontenc}
\RequirePackage{amsthm,amsmath}
\RequirePackage[authoryear,round]{natbib}
\RequirePackage[colorlinks,citecolor=blue,urlcolor=blue]{hyperref}
\RequirePackage[hmarginratio=1:1,top=32mm,columnsep=20pt]{geometry} % Document margins
\RequirePackage{textgreek}
\RequirePackage{bm}
\RequirePackage{stackengine}
\RequirePackage{amsfonts}
\RequirePackage{amssymb}
\RequirePackage{arydshln}
\RequirePackage[ruled,vlined]{algorithm2e}
\RequirePackage{graphicx} 
\RequirePackage{caption}
\RequirePackage{booktabs}
\RequirePackage{scrextend}
\RequirePackage{ulem}

\newtheorem{prop}{Proposition}
\newcommand{\tblcaption}[1]{\def\@captype{table}\caption{#1}}

\begin{document}

\begin{frontmatter}
\title{Structured penalized regression\\ for drug sensitivity prediction}
\runtitle{Structured penalized regression}
%\thankstext{T1}{Footnote to the title with the ``thankstext'' command.}
\footnotetext{{\it Address for correspondence: }Department of Biostatistics, Oslo Centre for Biostatistics and Epidemiology, Institute of Basic Medical Sciences, Faculty of Medicine, University of Oslo, P.O.Box 1122 Blindern 0317 Oslo, Norway. \\ E-mail: zhi.zhao@medisin.uio.no}
\begin{aug}
\author{Zhi Zhao}
\and
\author{Manuela Zucknick}

\affiliation{University of Oslo, Norway}

\end{aug}

\begin{abstract}
Large-scale {\it in vitro} drug sensitivity screens are an important tool in personalized oncology to predict the effectiveness of potential cancer drugs. The prediction of the sensitivity of cancer cell lines to a panel of drugs is a multivariate regression problem with high-dimensional heterogeneous multi-omics data as input data and with potentially strong correlations between the outcome variables which represent the sensitivity to the different drugs. We propose a joint penalized regression approach with structured penalty terms which allow us to utilize the correlation structure between drugs with group-lasso-type penalties and at the same time address the heterogeneity between omics data sources by introducing data-source-specific penalty factors to penalize different data sources differently. By combining integrative penalty factors (IPF) with tree-guided group lasso, we create the IPF-tree-lasso method. We present a unified framework to transform more general IPF-type methods to the original penalized method. Because the structured penalty terms have multiple parameters, we demonstrate how the interval-search Efficient Parameter Selection via Global Optimization (EPSGO) algorithm can be used to optimize multiple penalty parameters efficiently. Simulation studies show that IPF-tree-lasso can improve the prediction performance compared to other lasso-type methods, in particular for heterogenous data sources. Finally, we employ the new methods to analyse data from the Genomics of Drug Sensitivity in Cancer project.
\end{abstract}

\end{frontmatter}

\section{Introduction}\label{section:introduction}

With the advancements of high-throughput sequencing technologies, pharmacogenomics has been provided an emerging strategy for identifying personalized therapies from a mount of discovery drug candidates by using multi-omics data to characterize the whole biological system \citep{Hatzis2014, Bredel2004}. In personalised oncology in particular, large-scale in-vitro drug screens can be used to predict which drugs or drug combination will be effective for a patient given the particular molecular profile of the patient's tumour. 

In early publications \citep{Barretina2012, Garnett2012}, penalized linear regression methods, in particular the elastic net \citep{ZouHastie2005}, were used to demonstrate the potential of molecular data to predict the sensitivity of cancer cell lines to drugs. However, these papers only considered separate univariate models for each drug, and could thus not take the strong correlation structure between drugs due to similarities in drug function \citep{Ferdousi2017} into account. They also did not address heterogeneity between the different molecular data sources, for example expected differences in sparsity levels in the associations with the response variables \citep{Hasin2017}. In this work, we therefore propose and investigate a range of structured penalty terms in a multivariate linear regression setup, that allow us to take the known structure in the drug screen data better into account. 

In recent years, several groups and consortia have developed big datasets which include large-scale \textit{ex vivo} pharmacological profiling of cancer drugs on panels of cancer cell lines and the genomic profiling of these cancer cell lines \citep{Barretina2012, Daemen2013, Garnett2012, Greshock2010, Haverty2016}. The genomic data can for example consist of genome-wide measurements of mRNA expression, DNA copy numbers, DNA single-point and other mutations or CpG methylation of the cell lines taken at baseline, i.e. before treatment. They reflect different heterogeneous molecular profiles of the cancer cell lines with respect to effect sizes as level of sparsity in the effects (i.e. number of non-zero effects), correlations and collinearity, measurement scales and background noise \citep{Hasin2017}. The drug sensitivity profiles for some groups of drugs are expected to be correlated, due to their common targets and similar pharmacodynamic behaviours. 

To analyze such data, one straightforward method is to use (penalized) linear regression methods, regressing each drug on all molecular features in a linear manner, for example lasso \citep{Tibshirani1996} or the elastic net. Both methods can select a few relevant features with nonzero regression coefficient estimates from a large number of features. But they cannot address the heterogeneity of different molecular data sources. \citet{Boulesteix2017} introduced integrative $\ell_1$-penalized regression with penalty factors (IPF-lasso) to shrink the effects of features from different data sources with varying $\ell_1$-penalties, to reflect their different relative contributions. While lasso or IPF-lasso can be extended to multivariate regression to jointly model multiple drugs sensitivity, the correlation of drugs is not reflected in the penalization of regression coefficients. \citet{KimXing2012} proposed tree-guided group lasso (tree-lasso) to estimate structured sparsity of multiple response variables by assuming a hierarchical cluster structure in the response variables. Each cluster is likely to be influenced by some common features, for which the effects are similar between correlated responses.

In this article, we propose the IPF-tree-lasso which borrows the strength of varying penalty parameters from IPF-lasso and the cluster structure in multivariate regression from tree-lasso. Thus, IPF-tree-lasso can capture the different relative contributions of multiple omics input data sources and the group structure of correlated drug response variables. Since some cancer drugs might have similar mechanisms, for example the same target gene or signaling pathway, these drugs are likely to have correlated sensitivities. IPF-tree-lasso will select common relevant molecular features for these correlated drugs, and accordingly achieve similar regression coefficient estimates via a group-lasso-type penalty.

Elastic net \citep{ZouHastie2005} is also compared here, because it considers the grouping effect of correlated features and the $\ell_2$-penalty can improve the prediction performance over lasso. Additionally, we also formulate the integrative elastic net with penalty factors (IPF-elastic-net) model to provide a flexible extension of the elastic net with varying complexity parameters $\lambda$'s as well as varying parameters $\alpha$'s.

However, IPF-tree-lasso and IPF-elastic-net have more complicated penalty terms which might require new optimization algorithms. We use an augmented data matrix formulation, so that the original cyclical coordinate descent algorithm for lasso \citep{Friedman2010} and smoothing proximal gradient descent method proposed for tree-lasso \citep{KimXing2012} can be employed directly. As elastic net and IPF-type methods have multiple penalty parameters to be optimized, a standard grid search is computationally not efficient \citep{Jones1998}. Instead, we modify and employ the interval-search algorithm Efficient Parameter Selection via Global Optimization (EPSGO) proposed by \citet{FrohlichZell2005}. \Citet{Sill2014} implemented the EPSGO algorithm for elastic net models in the \texttt{R} package \texttt{c060}. We have adapted the \texttt{c060} package for efficient penalty parameters optimization for IPF-tree-lasso and IPF-elastic-net.

The introduction of structured penalty terms that are tailored to the specific structure of large-scale \textit{in vitro} drug sensitivity screening experiments can improve prediction performance, especially for groups of drugs that have the same mechanism of action and are therefore quite similar in their drug sensitivity profiles. But maybe even more importantly, the structured penalties also improve interpretation of the results, for example by highlighting to which groups of drugs (rather than only individual drugs) certain cell lines tend to show high drug sensitivity, or by allowing a comparison of the relative importance of the different omics data sources through their penalty factor ratios. In general, the joint analysis of all drugs and all cell lines from a drug sensitivity screen within a multivariate regression framework makes it easier to understand commonalities as well as differences in the drug sensitivity profiles, e.g., to understand which cell lines and tissue types show similar behaviour within groups of drugs with similar modes of action. Note that we extended all the methods discussed in this manuscript to allow the inclusion of un-penalised mandatory co-variates, in particular cancer tissue type.

The rest of the paper is organized as follows. In Section \ref{section:models}, we introduce the data setup and outline the consequences that this has for the construction of the structured penalty terms. We further briefly introduce all methods considered in the simulation studies and data application, from standard penalized regression methods to their newly developed extensions with structured penalty terms. The performance of the various methods is compared and evaluated in a simulation study described in Section \ref{section:simulation}, where the data are simulated to mimic typical large-scale drug screen scenarios. We finally perform a detailed analysis of data from the Genomics of Drug Sensitivity in Cancer project \citep{Garnett2012} in Section \ref{section:GDSC}, which includes a discussion of some highlights of the biological implications. Lastly, we discuss the main findings and conclude the paper in Section \ref{section:conclusion}.

\section{Structured penalties for multivariate regression}\label{section:models}

\begin{figure}
\centering\includegraphics[height=0.25\textwidth]{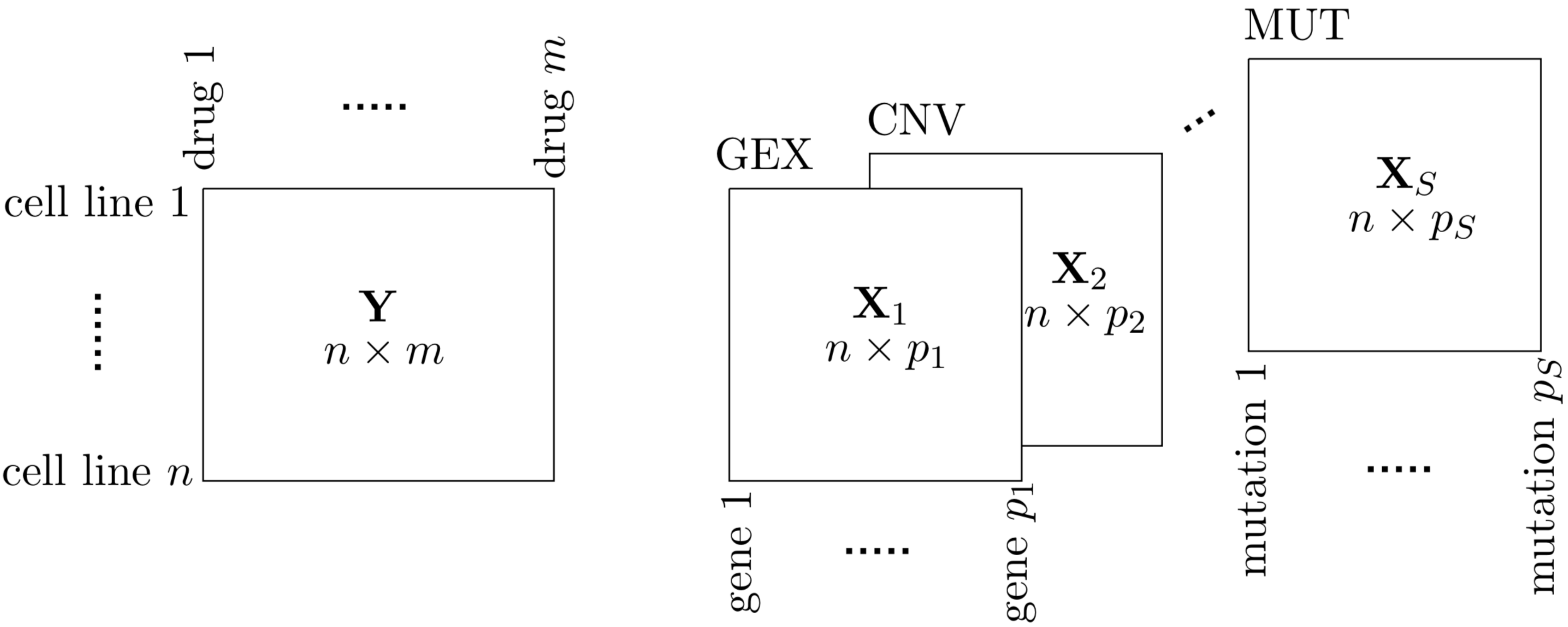}
\caption{Illustration of the response data and multiple data sources. The $\mathbf{Y}$ is measured from $n$ cell lines and $m$ drugs. For each cell line, gene expression (GEX) $\mathbf{X}_1$, copy number variation (CNV) $\mathbf{X}_2$, $\cdots$ and mutation (MUT) $\mathbf{X}_S$ are obtained which have $p_1$, $p_2$, $\cdots$ and $p_S$ features, respectively.}
\label{Fig1}
\end{figure}

The pharmacological data are collected for $n$ samples (e.g., cell lines or patients) and $m$ response variables (typically drug sensitivity). The response variables are denoted by $\mathbf{Y}=\{y_{ik}\}$ ($i \in [n] \equiv \{1,\cdots,n\}$, $k \in [m]$), where $y_{ik}$ means the response of the $i$th cell line treated with the $k$th drug as illustration in Figure \ref{Fig1}. The high-dimensional (i.e., multi-omics) data including $S$ data sources contain $p=\sum_{s=1}^S p_s$ features in total, and all $p$ features are available for all samples, denoted by $[\mathbf{X}_1,\cdots,\mathbf{X}_S]=\mathbf{X}=\{x_{ij}\}$ ($i\in [n]$, $j\in [p]$). The linear model mapping from high-dimensional data to multivariate responses is
\[
\mathbf{Y} =  \mathbf{1}_n\bm{\beta}_0^\top + \mathbf{XB} + \mathbf{E},\tag{2.1}\label{formula:linear}
\]
where $\mathbf{1}_n=(1,\cdots,1)^\top$ is an $n$-column vector, $\bm{\beta}_0=(\beta_{01},\cdots,\beta_{0m})^\top$ is the intercept vector corresponding to $m$ response variables, $\mathbf{B}$ is a $p \times m$ regression coefficients matrix, and $\mathbf{E}$ is a $n \times m$ noise matrix. $(\bm{\beta}_0, \mathbf{B})$ can be estimated by minimizing the sum of the residual sum of squares and a penalty function as following
\[
\underset{\bm{\beta}_0, \mathbf{B}}{\min}\left\{ \frac{1}{2mn}\|\mathbf{Y} - \mathbf{1}_n\bm{\beta}_0^\top - \mathbf{XB}\|_F^2 + \text{pen}(\mathbf{B}) \right\},\tag{2.2} \label{formula:penB}
\]
where $\|\cdotp\|_F$ is the Frobenius norm. 

\subsection{Lasso and elastic net} 

When $p$ is very large, especially $n \ll p$, and pressuming only a few true relevant features, the $\ell_1$-penalized regression, or lasso \citep{Tibshirani1996} is a standard method to identify those features by their nonzero coefficient estimates. The multivariate lasso uses the $\ell_1$-norm penalty function $\text{pen}(\mathbf{B})=\lambda\|\mathbf{B}\|_{\ell_1}$ in (\ref{formula:penB}), where $\|\mathbf{B}\|_{\ell_q} = (\sum_{j,k}|\beta_{jk}|^q)^{1/q}$ ($q \in [1, \infty)$) and $\lambda>0$ is the given penalty parameter that controls the strength of penalizing coefficients.

Another regularization method is elastic net \citep{ZouHastie2005}, which takes a bias-variance trade-off between lasso and the continuous shrinkage method ridge regression. The ridge penalty ($\ell_2$-penalty) tends to include or exclude strongly correlated features together. The penalty function of elastic net in (\ref{formula:penB}) is $\text{pen}(\mathbf{B})=\lambda(\alpha\|\mathbf{B}\|_{\ell_1} + \frac{1}{2}(1-\alpha)\|\mathbf{B}\|_{\ell_2}^2)$, where $\alpha \in [0,1]$ gives the compromise between ridge ($\alpha=0$) and lasso ($\alpha=1$).

\subsection{IPF-lasso and IPF-elastic-net} 

Lasso and elastic net penalize all coefficients of features by globally controlling penalty parameters $\lambda$ and $\alpha$. In order to distinguish the contributions of heterogeneous data sources, \citet{Boulesteix2017} proposed IPF-lasso to analyze multi-omics data. IPF-lasso allows varying penalty parameters to weight the norms of different sources' coefficients. For multivariate responses, the penalty function is $\text{pen}(\mathbf{B}) = \sum_s \lambda_s \|\mathbf{B}_s\|_{\ell_1}$ ($s\in [S]$), where $\lambda_s>0$, $\mathbf{B} = [\mathbf{B}_1 \Shortstack{. . .} \cdots \Shortstack{. . .} \mathbf{B}_S]$ stacks $\mathbf{B}_s$ by rows and $\mathbf{B}_s$ is the coefficients matrix corresponding to the $s$th data source. After transforming the $\mathbf{X}$ matrix as follows:
\begin{align*}
\mathbf{X}^\star &= 
	\begin{bmatrix}
	\mathbf{X}_1, \frac{\lambda_1}{\lambda_2}\mathbf{X}_2, \cdots, \frac{\lambda_1}{\lambda_S}\mathbf{X}_S
	\end{bmatrix}  \in \mathbb{R}^{n \times (p_1+ \cdots + p_S)}, \\
\mathbf{B}^\star&= 
	\begin{bmatrix}
	\mathbf{B}_1 \Shortstack{. . .} \frac{\lambda_2}{\lambda_1}\mathbf{B}_2 \Shortstack{. . .}\cdots \Shortstack{. . .} \frac{\lambda_S}{\lambda_1} \mathbf{B}_S
	\end{bmatrix} \in \mathbb{R}^{(p_1+ \cdots + p_S) \times m},
\end{align*}
then 
$\hat{\mathbf{B}}^\star = 
	\begin{bmatrix}
	\hat{\mathbf{B}}_1 \Shortstack{. . .} \frac{\lambda_2}{\lambda_1}\hat{\mathbf{B}}_2 \Shortstack{. . .}\cdots \Shortstack{. . .} \frac{\lambda_S}{\lambda_1} \hat{\mathbf{B}}_S
	\end{bmatrix}$
 where $\hat{\mathbf{B}} = [\hat{\mathbf{B}}_1 \Shortstack{. . .} \hat{\mathbf{B}}_2 \Shortstack{. . .}\cdots \Shortstack{. . .} \hat{\mathbf{B}}_S]$.

\begin{figure}
\centering\includegraphics[height=0.5\textwidth]{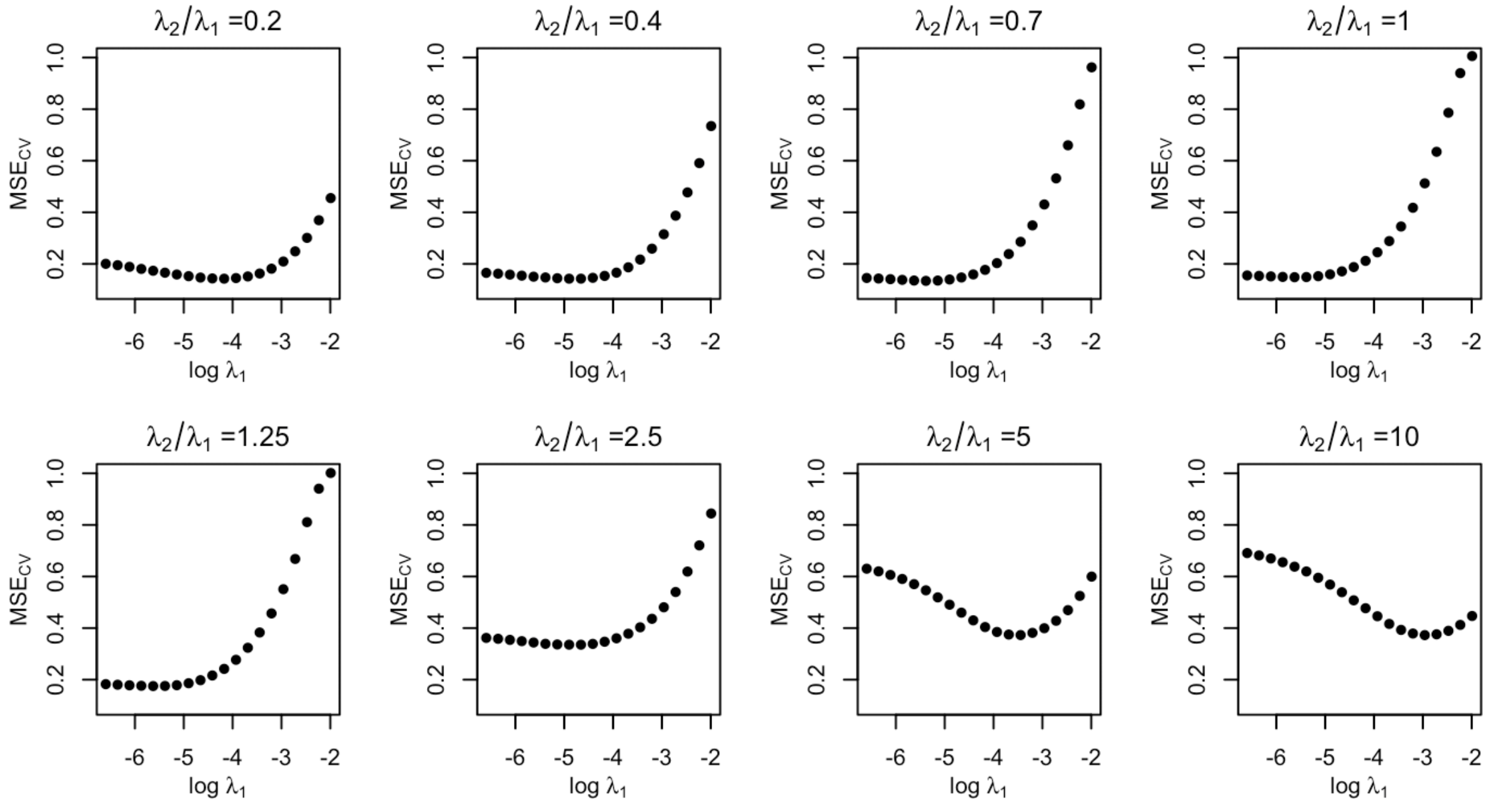}
\caption{Prediction performance of IPF-lasso for $(n,m,p_1,p_2)=(100,24,500,150)$ with different penalty parameter ratios $\lambda_2/\lambda_1$. It is exactly lasso when $\lambda_2/\lambda_1=1$. The details of the simulated data are as scenario 2 in Section \ref{section:sim_scenario}.}
\label{Fig2}
\end{figure}

When there are two data sources, i.e., $S=2$, Figure \ref{Fig2} shows that the prediction performance as mean squared error of cross validation, i.e., MSE$_{\text{CV}}$, is convex in $\lambda_1$ when fixing the two penalty parameters ratio $\lambda_2/\lambda_1$.

\cite{Boulesteix2017} mentioned the possible extension of IPF to the elastic net, but did not provide the formulation.  A simple integrative $\ell_1/\ell_2$-penalty method with penalty factors (sIPF-elastic-net) has $\text{pen}(\mathbf{B}) = \sum_s \lambda_s (\alpha \|\mathbf{B}_s\|_{\ell_1} + \frac{1}{2}(1-\alpha) \|\mathbf{B}_s\|_{\ell_2}^2)$ where $\lambda_s>0$ ($s\in [S]$) and $\alpha \in [0,1]$. Here, different data sources share the same parameter $\alpha$, which forces the same strength to shrink the coefficients of strongly correlated features towards each other. That is similar grouping effect as elastic net (see Supplementary S1 for more details). The fully flexible version of the IPF-elastic-net has penalty function
\[
\text{pen}(\mathbf{B})=\sum_s \lambda_s (\alpha_s \|\mathbf{B}_s\|_{\ell_1} + \frac{1}{2}(1-\alpha_s) \|\mathbf{B}_s\|_{\ell_2}^2), \tag{2.3}\label{formula:IPFEN}
\]
where $\lambda_s>0$, $\alpha_s \in [0,1]$ ($s\in [S]$). By defining the transformation
\begin{align*}
\mathbf{X}^\star&=
  \begin{bmatrix}
    \frac{1}{\alpha_1}\mathbf{X}_1  & \frac{\lambda_1}{\alpha_2\lambda_2} \mathbf{X}_2 & \dots &  \frac{\lambda_1}{\alpha_S\lambda_S} \mathbf{X}_S \\
    \frac{1}{\alpha_1}\sqrt{\frac{1}{2}\lambda_1(1-\alpha_1)} \mathbb{I}_{p_1} &  \mathbf{0}   & \dots & \mathbf{0} \\
    \mathbf{0} & \frac{\lambda_1}{\alpha_2\lambda_2} \sqrt{\frac{1}{2}\lambda_2(1-\alpha_2)} \mathbb{I}_{p_2} &   \dots & \mathbf{0}  \\
    \vdots & \vdots  & \ddots &  \vdots \\
    \mathbf{0} & \mathbf{0}  & \dots & \frac{\lambda_1}{\alpha_S\lambda_S} \sqrt{\frac{1}{2}\lambda_{S}(1-\alpha_{S})} \mathbb{I}_{p_{S}}   
  \end{bmatrix} \in \mathbb{R}^{(n+p) \times p}, \\
  \mathbf{Y}^\star &= [ \mathbf{Y} \vdots \mathbf{0} \vdots  \cdots \vdots \mathbf{0} ]  \in \mathbb{R}^{(n+p) \times m},  \\
\mathbf{B}^\star&= 
	\begin{bmatrix}
		\alpha_1\mathbf{B}_1 \Shortstack{. . .} \frac{\alpha_2\lambda_2}{\lambda_1}\mathbf{B}_2 \Shortstack{. . .}\cdots \Shortstack{. . .} \frac{\alpha_S\lambda_S}{\lambda_1} \mathbf{B}_S
	\end{bmatrix}  \in \mathbb{R}^{p \times m}, \\
 \lambda_1^\star&= \lambda_1/\{2m(n+\sum_s p_s)\}, \\
\alpha_s &\in (0,1], 
\end{align*}
it becomes a lasso problem
$$(\hat{\bm{\beta}}_0, \hat{\mathbf{B}}^\star)  = \underset{\bm{\beta}_0,\mathbf{B}^\star}{\arg\min}\left\{ \frac{1}{2mn}\|\mathbf{Y}^\star - \mathbf{1}_{n+p}\bm{\beta}_0^{\top} - \mathbf{X}^\star\mathbf{B}^\star\|_F^2 + \lambda_1^\star \|\mathbf{B}^\star\|_{\ell_1}\right\},$$
which only has the $\ell_1$-penalty term $\lambda_1^\star \|\mathbf{B}^\star\|_{\ell_1}$ given all $\alpha_s$ and $\lambda_s/\lambda_1$ ($s\in [S]$) are fixed.

\subsection{Tree-lasso and IPF-tree-lasso}\label{section:modelsTree}

\begin{figure}
\centering
\includegraphics[height=0.3\textwidth]{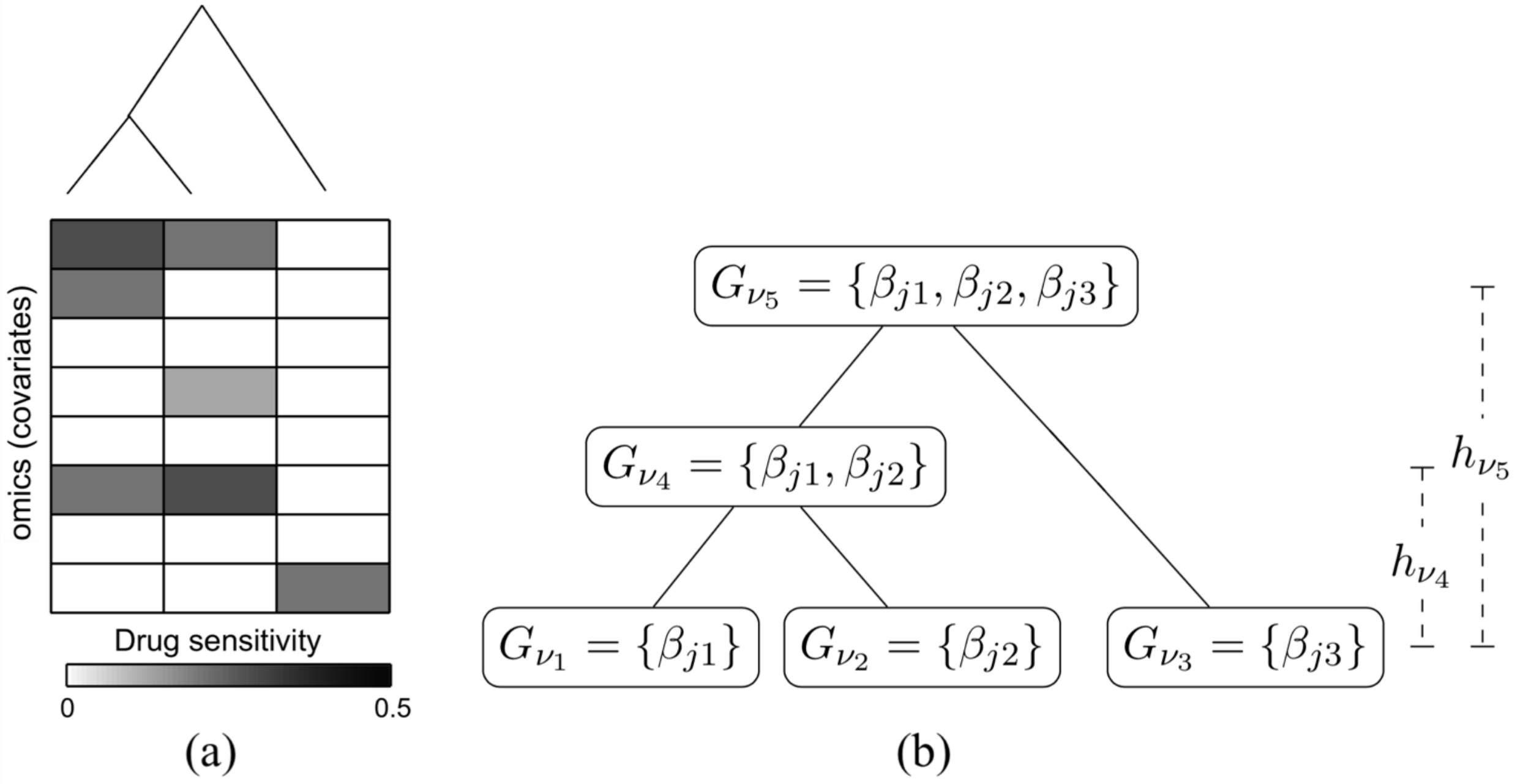}
\caption{An illustration of a tree lasso with three drugs (figure courtesy of \cite{KimXing2012}). (a): The sparse coefficients matrix is shown with white blocks for zeros and grey blocks for nonzero values. The hierarchical clustering tree above represents the correlation structure in drugs. The first two drugs corresponding to the first two columns are highly correlated and have two common influence covariates ($1$st and 6th rows). (b): Groups of coefficients associated with each node of the tree in panel (a) in the tree-lasso penalty. $\beta_{jk} (j \in [8], k \in [3])$ denotes the coefficient of the $k$th drug's $j$th variable.}
\label{Fig3}
\end{figure}

\citet{KimXing2012} proposed the tree-lasso method which uses a hierarchical tree structure over the response variables in a group-lasso based penalty function. As illustrated in Figure \ref{Fig3} we hypothesize that highly correlated response variables in each cluster are likely to be influenced by a common set of features. The hierarchical tree structure of multiple response variables can be represented as a tree $T$ with a set of vertices $V$ and groups $\{G_{\nu}: \nu \in V\}$. It can be given by prior knowledge of the pharmacokinetic properties of the drugs, or be learned from the data, for example by hierarchical clustering. Given the tree $T$ and groups $G_{\nu}$, the penalty function of the tree-lasso is defined as
\begin{align*}
\text{pen}(\mathbf{B}) &= \lambda \sum_{j=1}^p \sum_{\nu \in V} \omega_{\nu}\|\bm{\beta}_j^{G_{\nu}}\|_{\ell_2}  \\
&= \lambda \sum_{j=1}^p \sum_{\nu \in V_{\text{leaf}}} \|\bm{\beta}_j^{G_{\nu}}\|_{\ell_2} + \lambda \sum_{j=1}^p \sum_{\nu \in V_{\text{int}}} \omega_{\nu}\|\bm{\beta}_j^{G_{\nu}}\|_{\ell_2}\\
&= \lambda \sum_{j=1}^p \sum_{\nu \in V_{\text{leaf}}} |\bm{\beta}_j^{G_{\nu}}| + \lambda \sum_{j=1}^p \sum_{\nu \in V_{\text{int}}}\{h_{\nu} \cdot \sum_{c \in \text{Children}(\nu)} \omega_c\|\bm{\beta}_j^{G_c}\|_{\ell_2} + (1-h_{\nu})\|\bm{\beta}_j^{G_{\nu}}\|_{\ell_2}\},
\end{align*}
where $\bm{\beta}_j^{G_{\nu}} = \{\beta_{jk}: k \in G_{\nu} \}$ is the $j$th row of $\mathbf{B}$ associated with response variables in group $\nu$, $\omega_{\nu}$ is either the weight associated with the height $h_{\nu}$ of each internal node in tree $T$ or $\omega_{\nu}=1$ for the leaf node, $V_{\text{int}}$ and $V_{\text{leaf}}$ are the internal nodes and leaves of the tree, respectively. For example, consider a case with two drugs and a tree of three nodes that consists of two leaf nodes and one root node. It is illustrated as the following subtree of tree $T$ in Figure \ref{Fig3}, $V=\{\nu_1, \nu_2, \nu_4\}$, $\bm{\beta}_j^{G_{\nu_1}} = \{\beta_{j1}: j\in [p]\}$, $\bm{\beta}_j^{G_{\nu_2}} = \{\beta_{j2}: j\in [p]\}$, $\bm{\beta}_j^{G_{\nu_4}} = \{\beta_{jk}: j\in [p]; k\in \{1,2\}\}$. Then the penalty function for this tree is 
\[
\text{pen}(\mathbf{B})=\lambda\sum_{j=1}^p \{\underbrace{|\beta_{j1}|}_\text{leaf $G_{\nu_1}$}+\underbrace{|\beta_{j2}|}_\text{leaf $G_{\nu_2}$} + h_{\nu_4}(\underbrace{|\beta_{j1}|}_\text{leaf $G_{\nu_1}$}+\underbrace{|\beta_{j2}|}_\text{leaf $G_{\nu_2}$})+(1-h_{\nu_4})\underbrace{\sqrt{(\beta_{j1})^2+(\beta_{j2})^2}}_\text{internal $G_{\nu_4}$}\}. \tag{2.4}\label{formula:tree}
\] 
We can now define IPF-tree-lasso, where different $\lambda$'s are employed for different data sources. Its penalty function is defined as follows
\begin{equation}
\text{pen}(\mathbf{B}) = \sum_{s=1}^S \lambda_s\left(\sum_{j=1}^{p_s} \sum_{\nu \in V_{\text{int}}} \omega_{\nu}\|\bm{\beta}_{j,s}^{G_{\nu}}\|_{\ell_2} + \sum_{j=1}^{p_s} \sum_{\nu \in V_{\text{leaf}}} \|\bm{\beta}_{j,s}^{G_{\nu}}\|_{\ell_2}\right), \tag{2.5}\label{formula:IPFtree}
\end{equation}
where $\bm{\beta}_{j,s}$ is the $j$th row of coefficients corresponding to $s$th data source. The same transformation as IPF-lasso also makes IPF-tree-lasso to the equivalent tree-lasso problem (see Supplementary S2 for more details).

\begin{prop}\label{prop1}
Generally, the penalized objective function can be formulated as
\[
\underset{\bm{\beta}_0,\mathbf{B}}{\min} \left\{ \frac{1}{2mn}\|\mathbf{Y} - \mathbf{1}_n\bm{\beta}_0^\top - \mathbf{XB}\|_F^2 + \lambda \sum_{j=1}^p \sum_{g \in \mathcal{G}} w_g\|\mathcal{S}_j^g(\mathbf{B})\|_{\ell_{q_{j,g}}} \right\}, \tag{2.6}\label{formula:prop}
\]
where $\mathcal{S}_j^g(\mathbf{B})$ is the a submatrix of $\mathbf{B}$, $q_{j,g} \in [1,\infty]$, $\lambda$ is a tuning parameter, $\mathcal{G}$ contains pre-defined groups and all group weights $w_g$'s are known or can be pre-estimated. The submatrix $\mathcal{S}_j^g(\mathbf{B}) = \{\beta_{jk}: j \in M_g, k \in N_g\}$ is associated with specified set of rows $M_g \subseteq [p]$ and columns $N_g \subseteq [m]$. Let $\mathbf{X} = [\mathbf{X}_1, \dots, \mathbf{X}_S]$ with numbers of features $p_1, \dotsc, p_S$, respectively, $p=\sum_{s=1}^S p_s$ and $\mathbf{B} = [\mathbf{B}_1 \Shortstack{. . .} \cdots \Shortstack{. . .} \mathbf{B}_S]$. The corresponding IPF problem 
\begin{equation}
\underset{\bm{\beta}_0,\mathbf{B}}{\min} \left\{ \frac{1}{2mn}||\mathbf{Y} - \mathbf{1}_n\bm{\beta}_0^\top - \mathbf{XB}||_F^2 + \sum_{s=1}^S\sum_{j=1}^{p_s}\sum_{g\in \mathcal{G}} \lambda_s w_g\|\mathcal{S}_j^g(\mathbf{B}_s)\|_{\ell_{q_{j,g}}} \right\} \tag{2.7}\label{formula:propIPF}
\end{equation}
can be transformed into the equivalent original problem
$$\underset{\bm{\beta}_0,\mathbf{B}^\star}{\min}\left\{ \frac{1}{2mn}\|\mathbf{Y} - \mathbf{1}_n\bm{\beta}_0^\top - \mathbf{X}^\star\mathbf{B}^\star\|_F^2 + \lambda_1 \sum_{j=1}^p\sum_{g\in \mathcal{G}} w_g\|\mathcal{S}_j^g(\mathbf{B}^\star)\|_{\ell_{q_{j,g}}} \right\}, $$
where $\mathbf{X}^\star = [\mathbf{X}_1, \frac{\lambda_1}{\lambda_2}\mathbf{X}_2, \cdots, \frac{\lambda_1}{\lambda_S}\mathbf{X}_S]$, $\mathbf{B}^\star = [\mathbf{B}_1 \Shortstack{. . .} \frac{\lambda_2}{\lambda_1}\mathbf{B}_2 \Shortstack{. . .}\cdots \Shortstack{. . .} \frac{\lambda_S}{\lambda_1} \mathbf{B}_S]$ and $\mathcal{S}_j^g(\mathbf{B}^\star) = \{\beta_{jk}^\star: j \in M_g, k \in N_g\}$, if $\bigcup\limits M_g = [p]$ and $\bigcup\limits N_g = [m]$. If $\bigcup\limits M_g \subsetneqq [p]$ and (or) $\bigcup\limits N \subsetneqq [m]$, then the elements $\{\beta_{jk}^\star: j \in [p] \backslash \bigcup\limits M_g, k \in [m] \backslash \bigcup\limits N_g\}$ and the non-penalized features are remained in the Frobenius-norm loss function. The rows in $M_g$ and the columns in $N_g$ need not be contiguous, and $\mathcal{S}_j^g(\mathbf{B}_s)$ can be overlaping.
\end{prop}
The Proposition \ref{prop1} allows different norms for different submatrices in the penalty term. IPF-lasso and IPF-tree-lasso are special cases of (\ref{formula:propIPF}), that is,
\begin{itemize}
	\item {IPF-lasso: $\mathcal{G}=\{1\}$, $w_g=1$, $\mathcal{S}_j^g(\mathbf{B}_s)=\mathbf{B}_s$, $\ell_{q_{j,g}} = \ell_1$;} 
	\item {IPF-tree-lasso: $\mathcal{G}=\{V_{\text{leaf}},V_{\text{int}} \}$, $\mathcal{S}_j^g(\mathbf{B}_s)=\bm{\beta}_j^{\mathcal{G}_v}$, $\ell_{q_{j,g}}= \ell_2$.}
\end{itemize}
In (\ref{formula:prop}), $\|\mathcal{S}_j^g(\mathbf{B})\|_{\ell_{\infty}}=\text{sup}\{|\beta_{jk}|: j \in M_g, k \in N_g\}$ is likely to seek a common subset of a submatrix $\mathcal{S}_g(\mathbf{B})$, in which selected features will be relevant to multiple response variables simultaneously (Turlach and others, 2005). Since the penalty term in (\ref{formula:propIPF}) allows the overlaps of submatrices, it can contain more integrated penalty cases. For example, sparse-group lasso \citep{Simon2012} with $\ell_1$ and $\ell_2$ penalties actually allows the overlaps of coefficient groups (including singletons). \citet{Jacob2009} proposed group lasso allowing overlaps in combination with graphical lasso. \citet{Li2015} proposed the multivariate sparse group lasso, where an arbitrary group structure such as overlapping or nested or multilevel hierarchical structure is considered. All these methods can be extended into corresponding IPF-type methods and be solved by Proposition \ref{prop1}. However, (\ref{formula:propIPF}) does not include graphical lasso or fused lasso. If there is a non-identity transformation of the submatrix of $\mathbf{B}$ inside of the norm in the penalty term, the augmented matrix will be complicated. It also cannot be applied to the IPF-elastic-net because of the $\|\cdot\|_{\ell_2}^2$ penalty (\ref{formula:IPFEN}), unless $\frac{\lambda_1^2(1-\alpha_s)^2}{2\lambda_s\alpha_s^2}$ is a constant. 

\subsection{Implementation}

For the implementation, we give an initial decreasing $\lambda$ sequence, starting at $\lambda_{\text{max}}$ which shrinks all coefficients to zero \citep{Friedman2010}, for lasso, elastic net, IPF-lasso and sIPF-elastic-net. The multivariate responses $\mathbf{Y}$ are vectorized to $\text{vec}(\mathbf{Y})$. The interval-search algorithm Efficient Parameter Selection via Global Optimization (EPSGO) is applied to find the optimal $\alpha$ in elastic net type methods \citep{FrohlichZell2005, Sill2014}, as well as penalty parameters ratios in IPF-type methods. The EPSGO algorithm updates the tuning parameters through learning a Gaussian process model of the loss function surface from the points which have already been visited. Note that parameter tuning for IPF-elastic-net with varying $\alpha$'s remains challenging even when using the EPSGO algorithm due to the large number of parameters in a non-convex situation. For IPF-tree-lasso, $\lambda_1$ is optimized by exploring a given sequence of values, while the EPSGO algorithm is used to determine the optimal penalty parameters ratios $\lambda_s/\lambda_1$ ($s>1$). The IPF-tree-lasso is implemented based on the equivalent tree-lasso problem, which is more efficient than directly adapting the original tree-lasso algorithm to iterate coefficients of different data sources respectively (see Supplementary S3). The tree structure is pre-estimated from the response data  by hierarchical agglomerative clustering \citep{Golub1999} and only using the nodes with normalized heights larger than a pre-determined threshold $\rho^{\star}$ to ignore groups with weak correlations between response variables. Optimal penalty parameters are found by minimizing the MSE$_\text{CV}$ as loss function on the learning data, and independent validation data are used to obtain prediction mean squared errors (MSE$_{\text{val}}$) to evaluate prediction performance. 

\section{Simulations}\label{section:simulation}
\subsection{Simulation scenario}\label{section:sim_scenario}

We simulate data to demonstrate the prediction performance and variable selection performance of lasso, elastic net, tree-lasso and their corresponding IPF-type methods. We assume $n=100$ samples, multiple responses with $m=24$ drugs $\mathbf{Y}$, and two data sources of $p_1$ gene expression features $\mathbf{X}_1$ and $p_2$ mutation features $\mathbf{X}_2$ ($p_1>n, p_2>n$). The drug sensitivity responses are simulated according to the linear model
\begin{equation}
\mathbf{Y} = [\mathbf{X}_1, \mathbf{X}_2] 
	\begin{bmatrix}
	 \mathbf{B}_1 \\
	 \mathbf{B}_2
	 \end{bmatrix} + \mathbf{E}. \tag{3.1}
\end{equation}
To simulate the molecular data matrix $\mathbf{X}= [\mathbf{X}_1, \mathbf{X}_2]$, we first sample from a multivariate normal distribution with mean $\mathbf{0}$ and a nondiagonal $(p_1+p_2) \times (p_1+p_2)$ covariance matrix $\Sigma$ as \cite{Boulesteix2017} suggested, i.e.,
$\tilde{\mathbf{X}}=[\mathbf{X}_1, \tilde{\mathbf{X}}_2] \sim \mathcal{N}(\mathbf{0},\ \Sigma \otimes \mathbb{I}_n )$,
where 
\begin{equation*}
\Sigma
=
\left[
\begin{array}{ccc;{2pt/2pt}ccc}
    A_{p_1/b}(\sigma)  & \dots  & 0  & B_{p_1/b,p_2/b}(\sigma)  & \dots  & 0 \\
    \vdots & \ddots & \vdots  & \vdots & \ddots & \vdots \\
    0 & \dots  & A_{p_1/b}(\sigma)  & 0 & \dots  & B_{p_1/b,p_2/b}(\sigma)\\ \hdashline[2pt/2pt]
    B_{p_2/b,p_1/b}(\sigma)  & \dots  & 0  & A_{p_2/b}(\sigma)  & \dots  & 0 \\
    \vdots & \ddots & \vdots  & \vdots & \ddots & \vdots \\
    0 & \dots  & B_{p_2/b,p_1/b}(\sigma)  & 0 & \dots  & A_{p_2/b}(\sigma)
\end{array}
\right].
\end{equation*}
Here in each data source, we assume that there are $b$ latent groups of size $p_1/b$ or $p_2/b$, respectively. The covariance of any two features within the latent group is $\sigma$. In the covariance matrix $\Sigma$, blocks $A_{p_1/b}(\sigma)$ and $A_{p_2/b}(\sigma)$ capture the covariances of features within the first and second data source, respectively, and blocks $B_{p_1/b,p_2/b}(\sigma)$ and $B_{p_2/b,p_1/b}(\sigma)$ capture the covariances of the latent group features between the first and second data sources. The remaining parts of $\Sigma$ are zero. We set $\sigma=0.4$, $b=10$, and the variance of each feature is one. Further dichotomizing $\mathbf{X}_2 = \mathbf{1}_{\{\tilde{\mathbf{X}}_2 > 0\}}$ where $\mathbf{1}_{\{\cdot \}}$ is an indicator function and let $\mathbf{X}= [\mathbf{X}_1, \mathbf{X}_2]$. The second data source is dichotomized to simulate the common situation that one data source represents binary gene mutations. Considering that different data sources might have different numbers of features, we simulate two cases, $(p_1,p_2)=(150,150)$ and $(p_1, p_2)=(500,150)$. In (3.1), the noise term is $\mathbf{E}=\{ \epsilon_{ik} \}$ ($i \in [n]; k \in [m]$), where $\{\epsilon_{ik}\}\stackrel{iid}{\sim} \mathcal{N}(0,1)$.

\begin{figure}
\centering
\includegraphics[height=0.37\textwidth]{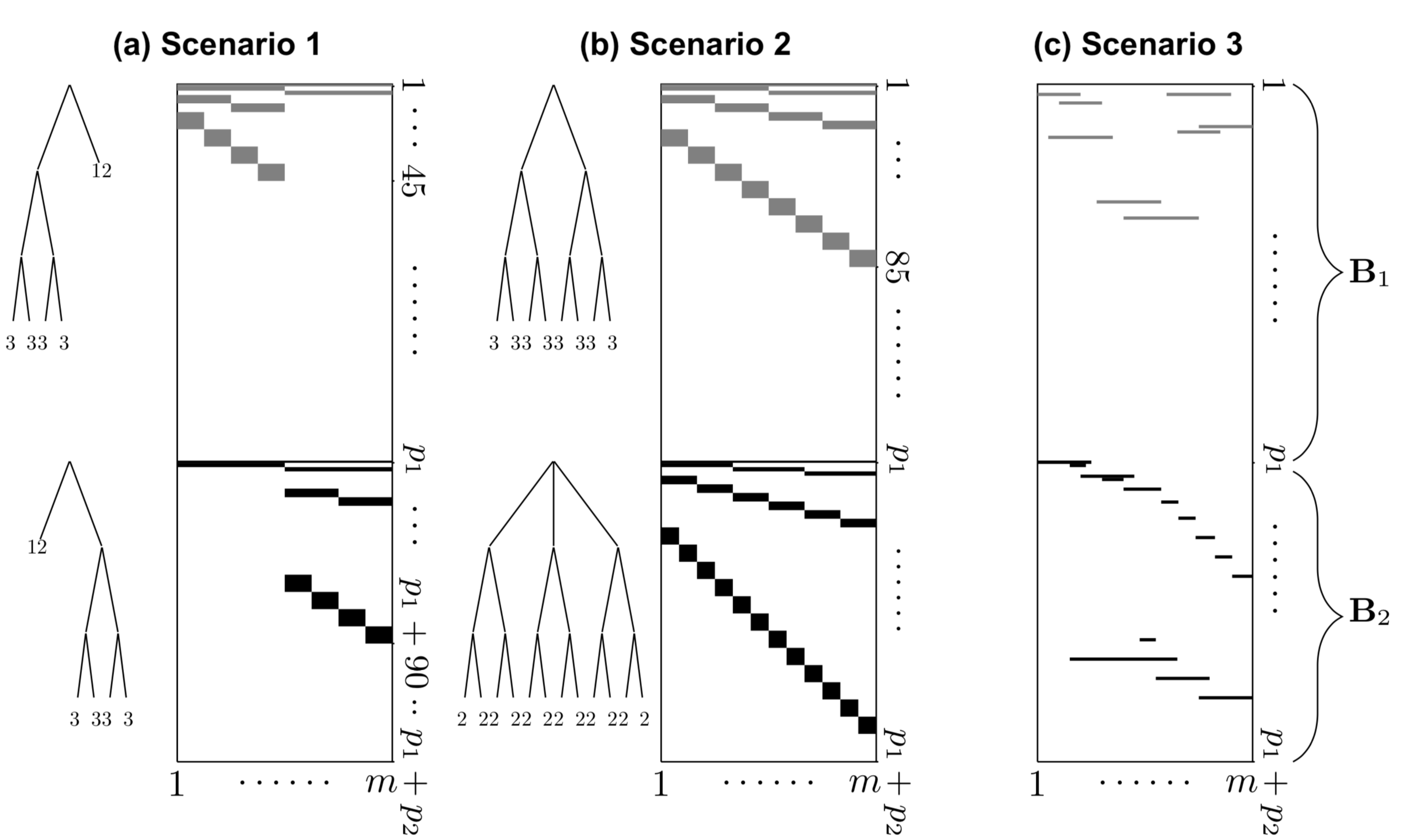}
\caption{Three simulation scenarios: structured $\mathbf{B}_1$ and $\mathbf{B}_2$ corresponding to hierarchical trees to generate hierarchically structured $\mathbf{Y}$ for scenarios (a) and (b). Black blocks indicate coefficient values of $0.6$, grey blocks $0.2$ and white areas $0$. The three scenarios have are 432, 720 and 216 nonzero coefficients, respectively. The two dendrograms on left of $\mathbf{B}$ in panel (a) or (b) show the hierarchical relationships of multivariate responses generated by $\mathbf{B}_1$ and $\mathbf{B}_2$, respectively. The number below each leaf node of the dendrogram denotes the number of response variables associated with that node (which in turn determines the widths of the gray/black blocks in the image plots of $\mathbf{B}$). The heights of the gray/black blocks represent the numbers of predictors associated with the response variables. Note, that in scenario (c) each of the grey/black blocks corresponds to two predictors. }
\label{Fig4}
\end{figure}

We assume that multiple responses can be grouped and the group relationships can be addressed by a hierarchical tree structure. In our intended applications, different molecular information may explain different group effects in the drug response variables. In the first simulation scenario, we design two different hierarchical tree effects from the two data sources, as illustrated in Figure \ref{Fig4}(a). The groups in the first 12 response variables are determined by the first data source, and the second data source determines the groups in the second 12 responses. The two hierarchical structures are generated by the two matrices $\mathbf{B}_1$ and $\mathbf{B}_2$ illustrated in Figure \ref{Fig4}(a). In a second simulation scenario the two data sources do not determine the drug groups separately, but instead in an overlapping manner. For this we design two very different hierarchical structures with $\mathbf{B}_1$ and $\mathbf{B}_2$, as illustrated in Figure \ref{Fig4}(b). The third simulation scenario has non-tree-structured $\mathbf{B}_1$ and $\mathbf{B}_2$ as shown in Figure \ref{Fig4}(c), in order to introduce a scenario with a more realistic structure in $\mathbf{B}$ that is not tailored to the tree-lasso-type methods. The patterns of $\mathbf{B}_1$ and $\mathbf{B}_2$ in Figure \ref{Fig4}(c) correspond to ``hotspots" generated as in the simulation study by \citet{Lewin2016}, where each response may be associated with a few features and each feature may be associated with multiple response variables.

To evaluate the prediction performance of different methods, we calculate MSE$_{\text{val}}$ from the simulated validation data $\mathbf{X}_{\text{val}}$ and $\mathbf{Y}_{\text{val}}$ which are simulated independently in an identical manner to $\mathbf{X}$ and $\mathbf{Y}$. Additionally, we compare the accuracy of variable selection performance, where we use the terms sensitivity to denote the percentage of nonzero coefficients accurately estimated as nonzeros and specificity to denote the percentage of zero coefficients accurately estimated as zeros. Algorithm 1 in Supplementary S4 summarizes the procedure of the simulation study.

\subsection{Simulation results and discussion}

\begin{figure}
\centering
\includegraphics[height=0.45\textwidth]{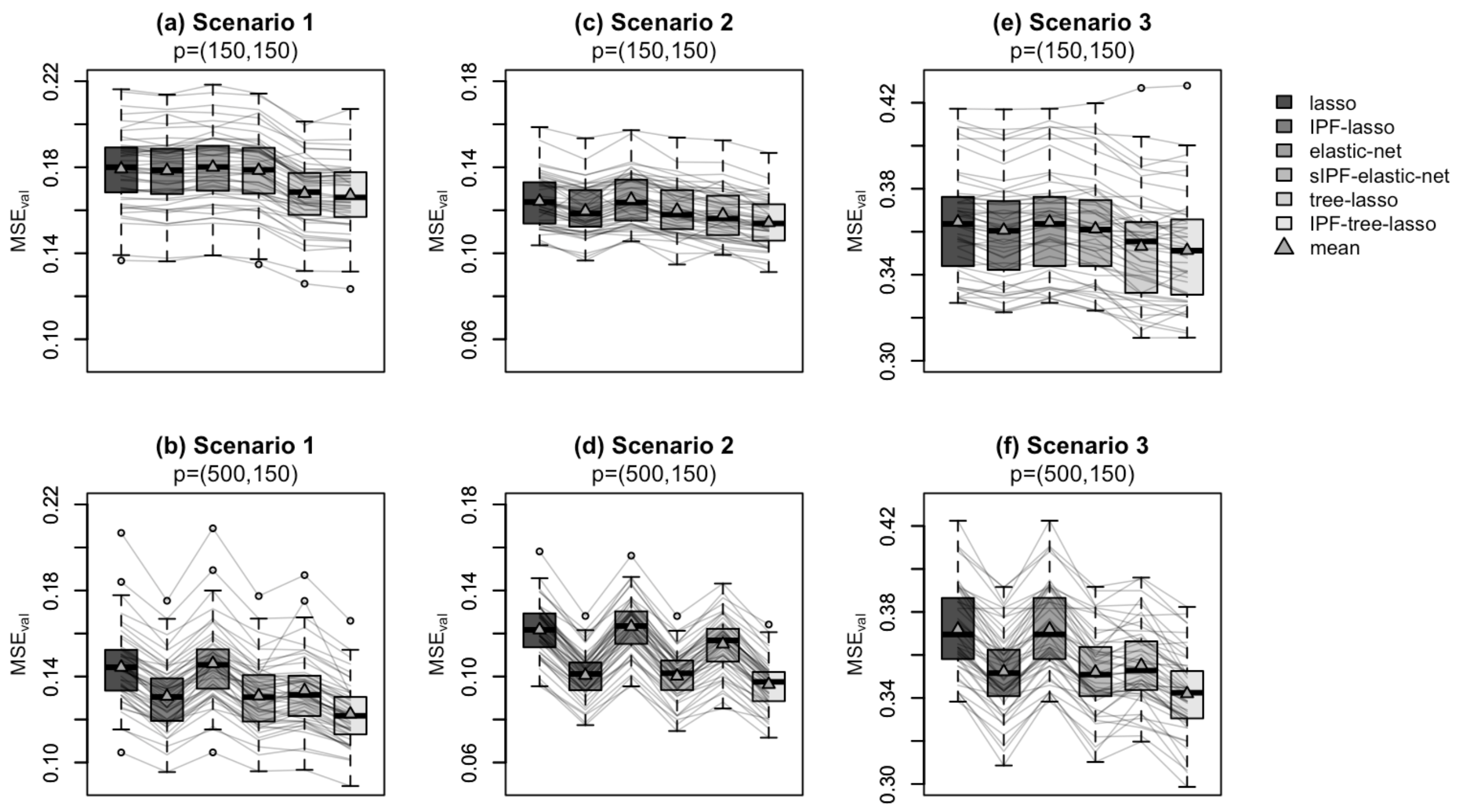}
\caption{Comparison of MSE$_{\text{val}}$ between different approaches in the three simulation scenarios with 50 simulations. The three panels of top are based on the same feature number of $\mathbf{X}_1$ and $\mathbf{X}_2$, $p_1=p_2=150$. The three panels of bottom are based on $p_1=500$ features of $\mathbf{X}_1$ and $p_2=150$ features of $\mathbf{X}_2$. Each box is drawn from 50 MSE$_{\text{val}}$ points corresponding to the 50 simulations. The grey dashed lines connect the performances of different models on the same data set.}
\label{Fig5}
\end{figure}

We run 50 simulations and show in Figure \ref{Fig5} the prediction performance of all methods in the different scenarios. When the two data sources have the same number of features (i.e., $p_1=p_2=150$) and similar trees for $\mathbf{B}_1$ and $\mathbf{B}_2$ as in scenario 1 (Figure \ref{Fig4}(a)), then lasso, IPF-lasso, elastic net and sIPF-elastic-net (i.e., simple IPF-elastic-net with common $\alpha$ for all data sources) have similar prediction performance in terms of MSE$_{\text{val}}$ (Figure \ref{Fig5}(a)) distribution. But tree-lasso and IPF-tree-lasso outperform other methods on average. However, when the two data sources are more different as in design scenario 2 (Figure \ref{Fig4}(b)), IPF-lasso and sIPF-elastic-net are slightly superior to lasso and elastic net in the situation of $p_1=p_2=150$ (Figure \ref{Fig5}(c)). IPF-tree-lasso further improves the prediction. In the non-tree-structured case, Figure \ref{Fig5}(e) (scenario 3), tree-lasso and IPF-tree-lasso still have better prediction.

When the two data sources have different feature numbers (i.e., $p_1=500$, $p_2=150$), IPF-type methods (IPF-lasso,  sIPF-elastic-net and IPF-tree-lasso) have lower MSE$_{\text{val}}$ than their corresponding non-IPF-type methods (lasso, elastic net and tree-lasso) respectively in all three scenarios (Figure \ref{Fig5}(b), (d) and (f)). When comparing scenario 2 to scenario 1 in the case $(p_1,p_2)=(500,150)$, IPF-type methods improve the average performance of prediction more in scenario 2. The grey dashed lines in Figure 5, which connect the performances of different models on the same data set, highlight that this is not only the case when comparing the methods with respect to their average performances (i.e. the mean MSE$_{\text{val}}$ values across all 50 simulated data sets), but that the rankings of the methods with respect to their MSE$_{\text{val}}$ values are very consistent across the individual data sets.

Table \ref{Table1} displays the accuracy of coefficients matrix estimation and variable selection for scenario 1. According to the averaged absolute errors of estimated coefficients, $\frac{1}{mp}\|\hat{\mathbf{B}}-\mathbf{B}\|_{\ell_1}$, all methods performe similarly if $(p_1,p_2)=(150,150)$ but the IPF-type methods perform slightly better for $(p_1,p_2)=(500,150)$. In both cases, $(p_1,p_2)=(150,150)$ and $(p_1,p_2)=(500,150)$, IPF-lasso,  sIPF-elastic-net, and IPF-tree-lasso select fewer features and achieve larger specificity than non-IPF-type methods, but they lose sensitivity. 

\begin{table}
\tblcaption{Accuracy of coefficients recovery by simulated hierarchical structured $\mathbf{Y}$ as scenario 1
\label{Table1}}
{%\small % 
\begin{tabular}{l rrrr}  
\toprule
& $\frac{1}{mp}\|\hat{\mathbf{B}}-\mathbf{B}\|_{\ell_1}$ & sensitivity & specificity & $VS^\dag$ \\
\cmidrule(r){2-5}
Model    & \multicolumn{4}{c}{$p_1=150$, \ \ $p_2=150$} \\
\midrule
Lasso                &   0.021 & 0.835	& 0.904	& 1008     \\
IPF-lasso           &  0.021	& 0.789	& 0.912	&  940    \\
Elastic-net           &  0.022 & 0.850	& 0.891	& 1103    \\
 sIPF-elastic-net  &  0.022	& 0.806	& 0.901	& 1016   \\
Tree-lasso           &  0.022	& 0.926	& 0.671	& 2627  \\
IPF-tree-lasso     &  0.022	& 0.917	& 0.680	& 2564  \\
\smallskip \\
& \multicolumn{4}{c}{$p_1=500$, \ \ $p_2=150$} \\
\cmidrule(r){2-5}
Lasso                    & 0.022  & 0.823	 & 0.943	 & 1219   \\
IPF-lasso              &  0.021 & 0.758	 & 0.959	 & 950  \\
Elastic-net             & 0.023   & 0.838	 & 0.936	 & 1329   \\
 sIPF-elastic-net    & 0.021  & 0.769	 & 0.955	 & 1011      \\
Tree-lasso            & 0.024    & 0.918	 & 0.755	 & 4121  \\
IPF-tree-lasso       &  0.022 & 0.903	 & 0.790	 & 3569   \\
\bottomrule
\multicolumn{5}{p{11cm}}{All results are the average of 50 simulations. Sensitivity denotes the percentage of nonzero coefficients estimated as nonzeros and specificity to denote the percentage of zero coefficients estimated as zeros. $^\dag$The variable selection index $VS = \sum_{j=1}^p\sum_{k=1}^m \mathbf{1}_{\{\hat{\beta}_{jk}^{(r)} \neq 0\}} $ indicates the number of selected features. Note that the selection is out of $p\times m$ features in total. } \\
\end{tabular}
}%
\end{table}

In the simulations, we mainly focus on simulating hierarchically structured drug sensitivity according to the specifically designed coefficient matrices as in Figure \ref{Fig4}(a) (scenario 1) and Figure \ref{Fig4}(b) (scenario 2). In both scenarios 1 and 2, the tree-lasso model improves performance. But even in the case of non-tree-structured $\mathbf{Y}$ (scenario 3) tree-lasso still outperforms lasso and elastic net in Figure \ref{Fig5}(e), because some of the correlation structures among response variables can be captured by a tree structure. For example, the nonzero blocks in $\mathbf{B}$ in Figure \ref{Fig4}(c) illustrate different response variables that are likely to be similar since these responses can be explained by the same features. \citet{KimXing2012} also showed that tree-lasso can take into account such correlations even when the tree structure is not fully realized. Compared to the top three panels of Figure \ref{Fig5}, the bottom three panels reflect a greater contribution to prediction performance of IPF-type methods. In all situations, IPF-tree-lasso achieves the best prediction, at least as good as tree-lasso when $p_1=p_2=150$, and the IPF-lasso and sIPF-elastic perform quite well when $(p_1,p_2)=(500,150)$. It is because IPF-tree-lasso does not only consider the correlations among responses, but also distinguishes the relative contributions of two data sources with varying penalty parameters $\lambda_1$ and $\lambda_2$. 

We monitored the running times of the different methods for one of the 50 simulations summarized in Figure \ref{Fig5}(a) on a 2.9 GHz Intel Core i5 MacBook Pro using three cores. With the \texttt{R} package \texttt{IPFStructPenalty} that we built, the six methods took 2.6sec, 52.9sec, 1.11min, 50.3sec, 3.96min, 41.30min, respectively (in the same order as shown in Figure \ref{Fig5}(a). The last method, which is IPF-tree-lasso, ran 7.06min with parallelization on a server with 10 cores. Note that with more than two data sources, the ESPSGO algorithm may need more careful specification of its optimization criteria to reach the global optimum, which would affect running times.

\section{Genomics of Drug Sensitivity in Cancer data analysis}\label{section:GDSC}

The Genomics of Drug Sensitivity in Cancer (GDSC) database \citep{Yang2013} was developed from a large-scale pharmacogenomic study, where a range of potential anticancer therapeutic compounds was tests on several hundred cancer cell lines. The drug sensitivity of a cell line, measured as half-maximal inhibitory concentration (IC$_{50}$), was estimated by fitting a Bayesian sigmoid model to dose-response curves obtained for a range of drug concentrations corresponding to the 72 hours' effect of drug treatment on cell viability, see \citet{Garnett2012} for the modelling details. We use the data from their archived files (\url{ftp://ftp.sanger.ac.uk/pub4/cancerrxgene/releases/release-5.0/}), where there are 97 cancer drugs tested on 498 cancer cell lines with complete availability of IC$_{50}$ measurements after excluding one cell line with one unrealistically small estimate for IC$_{50}$ $=1.43 \times 10^{-16}\mu M$. The cell lines represent tumor samples from 13 different tissue types. Cell lines are characterized by the following genomic data which are available as baseline measurements: genome-wide measurement of mRNA expression, copy numbers and DNA single point and other mutations. We preselect 2602 gene expression features with the largest variances over cell lines, which in total explain $50\%$ of the variation. In addition, we use copy number variation data for 426 genes and mutation data for 68 genes which are causally implicated in cancer according to the Cancer Gene Census (\url{https://cancer.sanger.ac.uk/census}).

We randomly select $80\%$ cell lines of each cancer tissue type for training data and the other $20\%$ as validation data. A multivariate regression model is fitted to the $80\%$ training data:
\begin{equation}
\mathbf{Y} =  \mathbf{1}_n\bm{\beta}_0^\top + \mathbf{X}_0\mathbf{B}_0 + \mathbf{XB} + \mathbf{E}, \tag{4.1}
\end{equation}
where $\mathbf{Y}$ is the $\log\text{IC}_{50}$, $\bm{\beta}_0$ is the intercept vector, $\mathbf{X}_0$ represents the cancer tissue types by dummy variables, and $\mathbf{X} = [\mathbf{X}_1, \mathbf{X}_2, \mathbf{X}_3]$ consists of the $\log$-transformed gene expression variables ($\mathbf{X}_1$), the counting copy number variables ($\mathbf{X}_2$) and binary mutation variables ($\mathbf{X}_3$). Let $\mathbf{B} = [\mathbf{B}_1 \Shortstack{. . .} \mathbf{B}_2 \Shortstack{. . .} \mathbf{B}_3]$ correspond to the coefficient matrices of the three data sources. Since tissue types are known to have large effects on drug sensitivity and $\mathbf{X}_0$ is low-dimensional, we do not penalize the coefficients $\mathbf{B}_0$ of cancer tissue types when fitting the model $(4.1)$. Supplementary S6 outlines the estimation of non-penalized coefficients in the tree-lasso model. The $20\%$ validation data are used to evaluate prediction performance by MSE$_{\text{val}}$. The implementation procedure of (4.1) by all methods is the same as the simulation studies (Algorithm 1 of Supplementary S4). Additionally, $R_{\text{val}}^2$ is computed as
$$R_{\text{val}}^2 \equiv 1-\frac{SSE_{\text{val}}}{SST_{\text{val}}} = 1-\frac{\|\mathbf{Y}_{\text{val}}-\mathbf{1}_n\hat{\bm{\beta}}_0^\top-\mathbf{X}_{\text{val}}\hat{\mathbf{B}}\|_F^2}{\|\mathbf{Y}_{\text{val}}-\mathbf{1}_n\bar{\mathbf{Y}}_{\text{val}}\|_F^2},$$
where $\mathbf{X}_{\text{val}}$ and $\mathbf{Y}_{\text{val}}$ are the 20\% validation data, $\bar{\mathbf{Y}}_{\text{val}}$ is a column vector with averaged $\log\text{IC}_{50}$ of each drug over the validation data samples. For comparison we also fit two low-dimensional linear regression models: (i) the ``NULL" model which only includes an intercept vector, and (ii) the ``OLS'' model which includes the intercept vector and dummy variables $\mathbf{X}_0$ representing the cancer tissue types.

To eliminate the uncertainty of splitting the data randomly into training$|$validation sets, Table \ref{Table2} reports the averaged results of 10 different random splits of the GDSC data. When only using the 13 tissue categories as predictors (``OLS" model), the prediction performance is very poor, MSE$_{\text{val}}$=3.199 and $R_{\text{val}}^2 = 0.036$ averaged over the 10 repetitions. However, including the genomic information improves the prediction in the lasso and elastic net models as shown in Table \ref{Table2}. By taking into account the hierarchical group relationship of 97 drugs (in tree-lasso and IPF-tree-lasso) and the heterogeneity of different data sources (in IPF-type methods), prediction performance can be improved further. IPF-tree-lasso performs best, with MSE$_{\text{val}}$=3.025 and $R_{\text{val}}^2 = 0.089$ averaged over the 10 repetitions. Nevertheless, as can be seen from $R_{\text{val}}^2$, all methods explain only a limited proportion of variation in the drug sensitivity data across all drugs. When looking at the drug-specific explained variation $R_{\text{val}}^2$ and MSE$_{\text{val}}$, the results differ widely between drugs. 

\begin{table}
\tblcaption{Prediction and the numbers of selected features in the GDSC data analysis
\label{Table2}}
{\small % 
\begin{tabular}{l rrrr}  
\toprule
\textbf{Method}  & NULL$^{\dag\dag}$ & Lasso & elastic net & Tree-lasso \\
\midrule
$VS^\star$$^{\dag}$ & -                      & 302+1+92$^{\sharp}$     & 315+1+93   & 21928+8149+1                                                \\
$\frac{1}{mp}VS^\star$$^{\ddag}$ & -                      & 0.1\%    & 0.1\%   & 10.0\%                                                \\
MSE$_\text{CV}$ (SD)$^{\S}$  &  3.360(0.027) & 3.200 (0.040) & 3.198 (0.039) & 3.138 (0.040)                                        \\   
MSE$_{\text{val}}$ (SD)  		     &   3.368 (0.107) & 3.151 (0.077) & 3.149 (0.077) & 3.069 (0.079)                                                        \\
$R_{\text{val}}^2$$^{\P}$ (SD)  	     &   -0.014 (0.008)  & 0.051 (0.012)     & 0.052 (0.014)     & 0.076 (0.019)                                 \\

\\
& OLS$^{\ddag\ddag}$  & IPF-lasso &  sIPF-elastic-net & IPF-tree-lasso \\
\cmidrule(r){2-5}              
$VS^\star$   & -                     & 774+11+74 & 252394+41322+6596 & 30567+515+452                                        \\
$\frac{1}{mp}VS^\star$ & -                      & 0.3\%    & 100.0\%   & 10.5\%                                                \\
MSE$_\text{CV}$ (SD)    & 3.013 (0.016) & 3.182 (0.037) & 3.179 (0.036) & 3.068 (0.035)                 \\   
MSE$_{\text{val}}$ (SD)                   & 3.199 (0.074) & 3.134 (0.078) & 3.130 (0.076) & 3.025 (0.074)                  \\
$R_{\text{val}}^2$ (SD)                  & 0.036 (0.016)     & 0.056 (0.014)     & 0.057 (0.015)      & 0.089 (0.018)                      \\

\bottomrule
\multicolumn{5}{p{13cm}}{$^{\dag}$The variable selection index $VS^\star = \sum_{j=1}^p\sum_{k=1}^m \left\{ \left(\sum_{r=1}^{10}\mathbf{1}_{\{\hat{\beta}_{jk}^{(r)} \neq 0\}} \right) \geq 2 \right\} $ indicates the features selected at least twice over the $10$ repetitions. $^{\ddag}$$\frac{1}{mp}VS^\star$ is the proportion of the number of selected features. $^{\S}$MSE$_{\text{cv}}$ is the mean squared error of 5-fold cross-validation based on the $80\%$ training cell lines, and MSE$_{\text{val}}$ is the predicted mean squared error based on the $20\%$ testing cell lines. Both of them are the average of the $10$ repetitions. The SD is the standard derivation over the $10$ repetitions. The SD should be considered with caution, since the 10 repetitions are correlated. $^{\P}$The range of $R_{\text{val}}^2$ is $(-\infty,1]$. $^{\dag\dag}$The NULL model is $\mathbf{Y}=\mathbf{1}_n\bm{\beta}_0^\top+\mathbf{E}$. $^{\ddag\ddag}$The OLS model only includes the $13$ cancer tissue types information as predictors, i.e., $\mathbf{Y}=\mathbf{1}_n\bm{\beta}_0^\top + \mathbf{X}_0\mathbf{B}_0+\mathbf{E}$ where $(\hat{\bm{\beta}}_0,\hat{\mathbf{B}}_0)$ are OLS estimates. $^{\sharp}$The number of estimated nonzero coefficients, \#\{$\hat{\beta}_{jk} \neq 0: j \in [p], k \in [m]$\}, corresponds to the mRNA expression, copy numbers and mutation, which are selected at least twice in the $10$ repetitions over all drugs. Note that the selection is out of $p\times m=300\ 312$ features in total. } \\
\end{tabular}
}%
\end{table}

\begin{figure}
\centering
\makebox[\textwidth][c]{\includegraphics[height=0.58\textwidth]{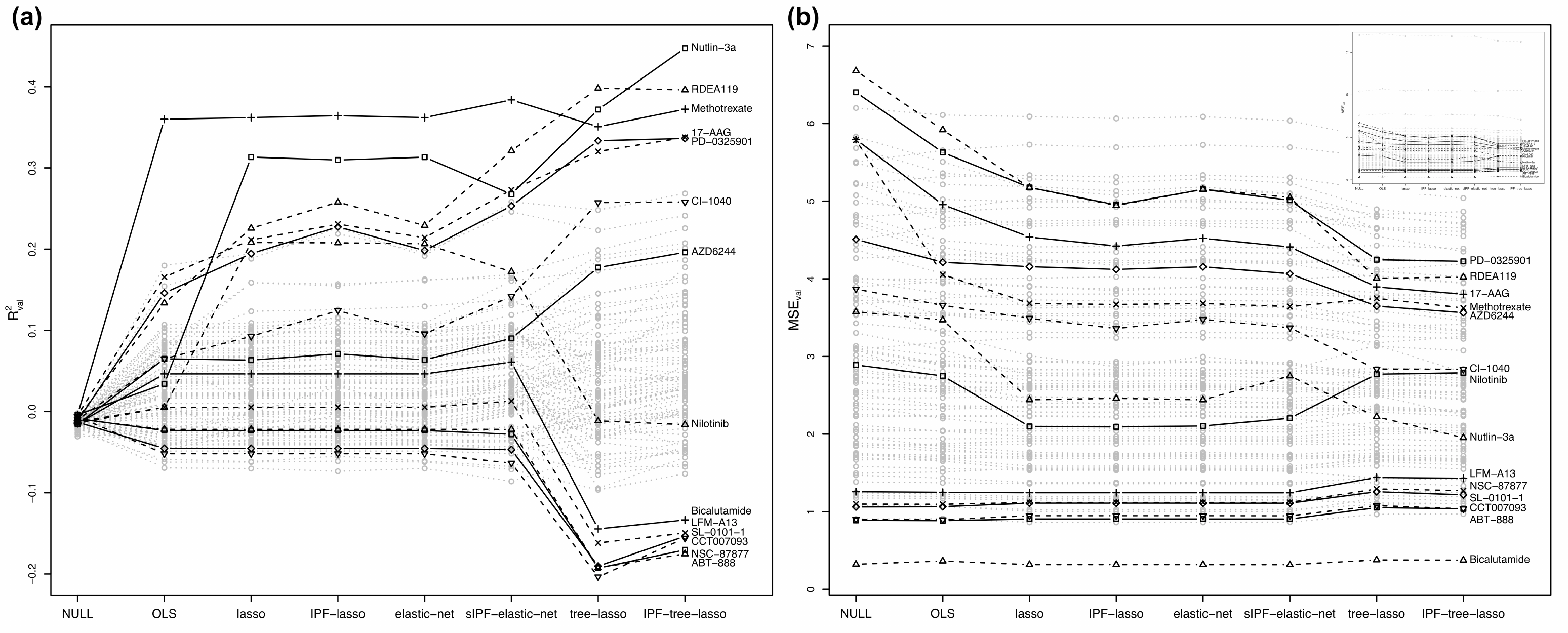}}
\caption{$R_{\text{val}}^2$ and MSE$_{\text{val}}$ for individual drugs by each method. The results are the average of 10 repetitions and based on the final models of each repetition. The top-right of (b) shows the full y-axis with all drugs.}
\label{Fig6}
\end{figure}

For example, the IPF-tree-lasso model can explain $44.8\%$ of the variation between cell lines in their sensitivity to Nutlin-$3\alpha$ (Figure \ref{Fig6}(a)). Nutlin-$3\alpha$ has been reported to be effective in cancers across a wide range of tissue types due to its general mechanism of action, which involves the p53-pathway that is affected in up to $50\%$ of cancers in nearly all cancer types \citep[e.g.,][]{Olivier2002,Luo2013,Trino2016,Drakos2011,Hui2018}. Nutlin-$3\alpha$ therefore reaches a relatively large proportion of cancers, which in addition can be characterized well by their molecular profiling, in particular by their TP53 status and MDM2 expression. Consequently, improving the estimates for these molecular features, e.g., by borrowing information across related p53-targeting drugs as can be achieved by tree-lasso and IPF-tree-lasso, is particularly helpful for improving overall prediction performance of Nutlin-$3\alpha$. When we further split the prediction performance of IPF-tree-lasso for Nutlin-$3\alpha$ by calculating the $R_{\text{val}}^2$ separately for each cancer tissue type (Table S7.1 of Supplementary S7), we observe high explained variation ($R_{\text{val}}^2>0.180$) in the digestive system, urogenital system, blood, kidney, nervous system, skin, soft tissue, aerodigestive tract, lung, pancreas and bone tissue. 

In addition to Nutlin-$3\alpha$, drugs RDEA-119, 17-AAG, PD-0325901, CI-1040 and AZD6244 also reach much higher $R_{\text{val}}^2$ for tree-lasso and IPF-tree-lasso than for the other methods (Figure \ref{Fig6}(a)). This indicates that these drugs benefit particularly well from the borrowing of information across related drugs. In fact, drugs RDEA-119, PD-0325901, CI-1040 and AZD6244 are all MEK inhibitors. These drugs with the same target have highly correlated drug sensitivities and are clustered together in the hierarchical clustering used to determine the tree structure for tree-lasso and IPF-tree-lasso (Figure S7.1 of Supplementary S7). Although the drug 17-AAG is a Hsp90 inhibitor, \citet{Calabrese2003} showed that down-regulation of MEK1 activity markedly reduces the sensitivity of medulloblastoma, breast, and ovarian cancer cells to 17-AAG; therefore 17-AAG can also benefit from the borrowing of information in the tree-lasso methods, since it clusters quite closely together with the above-mentioned MEK inhibitors.

As for the drug Methotrexate, all prediction models except the ``NULL'' model perform similarly well. Methotrexate is a cytotoxic chemotherapeutic drug which blocks the enzyme dihydrofolate reductase which in turn results in cell death \citep{Goodsell1999}. It is a very general mechanism of action and not targeted to specific genes or pathways, so consequently the omics characterisation of the cell lines does not contribute further to the prediction performance for this drug. Even when only considering the tissue types without genomic information, the ``OLS'' model for Methotrexate already reaches very good prediction performance ($R_{\text{val}}^2 = 0.360$).

Note, that for a few drugs, tree-lasso and IPF-tree-lasso do not perform well. In particular, the cell lines' sensitivity to Nilotinib is on average predicted relatively well by lasso, elastic-net, IPF-lasso and sIPF-elastic-net ($R_{\text{val}}^2>0.170$), but cannot be predicted by tree-lasso and IPF-tree-lasso (Figure \ref{Fig6}(a)) where $R_{\text{val}}^2<0.002$. When comparing the prediction performance of Nilotinib for individual cancer tissue types, we have found that only blood cell lines have much smaller MSE$_{\text{val}}$ for lasso, elastic-net, IPF-lasso and sIPF-elastic-net compared to the tree-lasso-type methods. This may be because the mode of action of Nilotinib is specific to chronic myeloid leukemia (a blood cancer), where it blocks a tyrosine kinase protein coded by the fusion gene BCR-ABL which is found in most patients with chronic myeloid leukemia \citep{BlayVonMehren2011}. Therefore, the presence of the BCR-ABL fusion gene is a strong predictor for sensitivity to Nilotinib in blood cancer cell lines. In fact, while there are only three blood cancer cell lines with BCR-ABL mutation in this data set, all three show a very good sensitivity to Nilotinib with extremely low IC$_{50}$ values (Figure S7.2 of Supplementary S7).

All lasso-type methods select the gene BCR-ABL for Nilotinib for all repetitions and achieve good prediction for the three BCR-ABL mutated blood cell lines. But tree-lasso neither selects the BCR-ABL mutation variable nor predicts drug sensitivity of the blood cell lines well for Nilotinib, and although it is selected for Nilotinib by IPF-tree-lasso, the absolute value of its coefficient estimate is very small. This is because the IC$_{50}$'s  of drugs Nilotinib and Axitinib have strong correlation (Pearson correlation coefficient 0.57, Figure S7.3 of Supplementary S7), and therefore the tree-lasso penalties shrink the coefficients corresponding to these two drugs together which results in more similar predictions. While the three BCR-ABL-mutated blood cell lines also show strong sensitivity to Axitinib, the corresponding IC$_{50}$ are not as extreme as for Nilotinib, so that the shrinkage of the regression coefficients of Nilotinib and Axitinib towards each other results in a worse prediction for Nilotinib for these three extreme cell lines. This is reflected in the overall $R_{\text{val}}^2$, which is strongly influenced by these three extreme data points.

Other drugs including Bicalutamide, LFM-A13, SL-0101-1, CCT007093, NSC-87877 and ABT-888 show small $R_{\text{val}}^2$ values ($R_{\text{val}}^2 < 0.050$) for all methods (Figure \ref{Fig6}(a)). All these drugs have very small MSE$_{\text{val}}$ values, even for the NULL model (Figure \ref{Fig6}(b)), which indicates low heterogeneity in the $\log \text{IC}_{50}$ values for these drugs. In short, there is not much variability between cell lines that could be explained by the molecular data or tissue type information available.
 
In Table \ref{Table2}, $VS^\star = \sum_{j=1}^p\sum_{k=1}^m \left\{ \left(\sum_{r=1}^{10}\mathbf{1}_{\{\hat{\beta}_{jk}^{(r)} \neq 0\}} \right) \geq 2 \right\} $ indicates the number of features selected at least twice over the ten repetitions\footnote{Accordingly, throughout this paragraph we refer to a feature as being selected, if it has an estimated non-zero regression coefficient in at least two out of ten repetitions.}. We note that lasso, elastic net and IPF-lasso perform very sparse variable selection with fewer than 860 out of $p\times m=300\ 312$ features estimated  to be nonzero more than once over ten repetitions, which corresponds to $0.3\%$ of all features. Although sIPF-elastic-net has slightly better average prediction performance than standard lasso and elastic net, it results in an almost dense coefficient matrix with $\alpha=0.39$ (where $\alpha$ is the averaged optimal value over the 10 repetitions) compared to the elastic net with $\alpha=0.77$. Both tree-lasso-type methods are sparser than sIPF-elastic-net with $VS^\star = 30\ 078$ for tree-lasso and $VS^\star = 31\ 534$ for IPF-tree-lasso (corresponding to $10.0\%$ and $10.5\%$ of all features, respectively). Tree-lasso selects many more copy number features than IPF-tree-lasso, but only one mutation feature: TP53 for drug Nutlin-$3\alpha$. In contrast, IPF-tree-lasso selects 537 associated mutated gene features. In particular, the following drugs with $R_{\text{val}}^2>0.100$ select the following mutation features corresponding to the drugs' respective targets: B-Raf for drugs PLX4720 and SB590885, EGFR/ErbB for drug BIBW2992 (Afatinib), BCR-ABL for Nilotinib and TP53 for Nutlin-$3\alpha$. In addition, drugs RDEA119, 17-AAG, PD-0325901, CI-1040 and AZD6244 all select genes in their target Ras/Raf/MEK/ERK signalling pathway by IPF-tree-lasso. All target genes selected by different methods for drugs with $R_{\text{val}}^2>0.100$ are shown in Table S7.2 of Supplementary S7.

\section{Conclusion}\label{section:conclusion}

In this study, we used multi-omics data to jointly predict sensitivity of cancer cell lines to multiple cancer drugs. It was our goal that the penalized regression should take into account both, the heterogeneity in the multi-omics data as well as (hierarchical) relationships between drugs. We extended the tree-lasso to IPF-tree-lasso, which can achieve the two purposes simultaneously. In addition, based on IPF-lasso \citep{Boulesteix2017}, we formulated the IPF-elastic-net, which is an option to tune the varying penalty parameters $\lambda_s$ and $\alpha_s$ in (\ref{formula:IPFEN}) separately for each data source $s \in [S]$, thus allowing differing degrees of both, sparsity and grouping effect, in the different data sources. If one has prior knowledge on the sparsity of the different data sources, IPF-elastic-net can specify the penalty terms for some data sources as lasso penalties for instance, i.e. specify those $\alpha_s$ values as 1. Finally, we provide a unified framework for the transformation of a large class of general IPF-type penalized problems into the equivalent original penalized problems in proposition \ref{prop1}. Furthermore, we demonstrated how the interval-search algorithm EPSGO \citep{FrohlichZell2005} can be used to optimize multiple penalty parameters efficiently. 

To capture the heterogeneity of different features, \citet{Bergersen2011} proposed the weighted lasso to use external information to generate penalty weights, e.g., to use external copy number data to provide weights for gene expression features. However, in the IPF-type methods all data contribute to the outcomes directly. \citet{VandeWiel2016} developed adaptive group-regularized ridge regression (``GRridge") to use related co-data (e.g., annotation or external $p$ values) to derive group-specific penalties by empirical Bayes estimation; only one global penalty parameter needs to be optimized, e.g., by cross-validation. \citet{DondelingerMukherjee2018} proposed joint lasso to penalize subgroup-specific (i.e., cancer tissue) coefficients differently with an additional fusion penalty term, but they used the same penalty parameter value for all high-dimensional features. \citet{Klau2018} presented priority-lasso to construct blocks of multi-omics data sources and regress on each data source sequentially. It needs a priority sequence for regressing the multiple data sources. Thus the resulting variable selection favors those with higher priority. Although the weighted lasso, GRidge and priority-lasso were developed for univariate response, they can be extended to independent multivariate responses. \citet{Wu2019} selectively reviewed multi-level omics data integration methods, but focused on univariate outcomes.

In our analysis of the GDSC data set, the overall low $R_{\text{val}}^2$ highlights the limitations in using only tissue type and genomic information to predict drug sensitivity. \citet{Wang2017} and \citet{Ali2018} showed that proteomes of human cancer cell lines are more representative of primary tumors than genomic profiles alone and might thus improve the prediction of drug sensitivity. \citet{ChamblissChan2016} recommended to integrate pharmacoproteomics and pharmacogenomics profiles of the tumor samples to help identify the right therapeutic regimen. Throughout this article, all data are assumed to be complete. If data are missing completely at random, the EM algorithm can be used to impute missing values, for example the EMLasso can be adapted to our approaches \citep{Sabbe2013}.  

One disadvantage of IPF-type methods is that they cannot address known associations between features in the different data sources, e.g., between gene expression and mutation status of the same gene. However, the  sIPF-elastic-net employs a common $\alpha$ for all data sources, which is likely to select the strongly correlated features over all data sources together if $\alpha$ is small. As for IPF-tree-lasso, specifying similar weights for coefficients of associated internal nodes in (\ref{formula:IPFtree}) across different data sources may select correlated features over multiple data sources. Furthermore, the inclusion of biological pathways of genes related to the cancer may improve the biological interpretation and prediction of drug sensitivity as well. \cite{LiLi2008} and \citet{Lee2016} proposed pathway-based approaches to identify biologically relevant pathways related to interesting phenotype. The tree-lasso or IPF-tree-lasso can also be extended to include the pathway-group structure over features in the penalty term besides the hierarchical structure over response variables.

\appendix

\section*{Acknowledgements}

The first author is supported by the Faculty of Medicine, University of Oslo, Norway. This work is associated with the project BIG INSIGHT: 237718 funded by the Research Council of Norway. The authors thank Professor Arnoldo Frigessi for discussions. {\it Conflict of Interest}: None declared.

\section*{Supplementary Material}

Software in the form of \texttt{R} package \texttt{IPFStructPenalty} is available on
 \url{https://github.com/zhizuio/IPFStructPenalty}.  All simulations and data analyses were performed in R version v3.5. The simulation studies can be reproduced with the file \texttt{Sim\_reg.R} to generate the simulated data, analyze the data and generate the output figures and tables shown in this article. The data analysis can be reproduced with the file \texttt{GDSC\_reg.R}, which provides information on data retrieval, the instruction how to generate the GDSC data set used in this article and the file  \texttt{GDSC\_reg.R} which can be used to fit the different models. The simulation and data analysis scripts (rewritten for serial execution) are provided on \url{https: //github.com/zhizuio/IPFStructPenalty}. Note that, if the reader is interested in replicating the entire study, parallel execution is preferred, e.g., in an HPC environment, because of the high computational cost. Other supplementary materials including proofs, extra algorithms and results are available online.

%\begin{supplement}
%\sname{Supplement A}\label{suppA}
%\stitle{Title of the Supplement A}
%\slink[url]{http://www.e-publications.org/ims/support/dowload/imsart-ims.zip}
%\sdescription{Dum esset rex in
%accubitu suo, nardus mea dedit odorem suavitatis. Quoniam confortavit
%seras portarum tuarum, benedixit filiis tuis in te. Qui posuit fines tuos}
%\end{supplement}

\clearpage

\clearpage

\end{document}